\documentclass[a4paper]{scrartcl}
\input {notes.sty}

\newcommand{\soten}{$SO(10)$-inspired}
\newcommand{\tre}{\theta_{13}}
\newcommand{\tatm}{\theta_{23}}
\newcommand{\mee}{m_{ee}}
\newcommand{\preex}{N_{B-L}^{preex}}
\newcommand{\lept}{N_{B-L}^{lept}}
\renewcommand{\d}{$\delta$}
\renewcommand{\deg}{\degree}

\begin{document}

%===============================================================================
\title{Strong thermal Leptogenesis: an exploded view of the low energy neutrino parameters in the $SO(10)$-inspired model\footnote{Based on \cite{DiBari:2013qja}. Proceedings for the XXX-th International Workshop on High Energy Physics ``Particle and Astroparticle Physics, Gravitation and Cosmology: Predictions, Observations and New Projects'', June 23-27 2014, Protvino, Russia.}}
\author{Luca Marzola\\
\\
\small{\emph{Laboratory of Theoretical Physics, Institute of Physics, University of Tartu,}}\\
\small{ 
\emph{Ravila 14c, Tartu, Tartu county 50411, Estonia.}}\\
\small{\texttt{luca.marzola@ut.ee}}}
\date{}
\maketitle

%===============================================================================

\begin{abstract}
Leptogenesis is an attractive scenario in which neutrino masses and baryon asymmetry of the Universe are explained together under a minimal set of assumptions. 

After formulating the problem of initial conditions and introducing the strong thermal leptogenesis conditions as solution, we show that, within the framework provided by the \soten~model of leptogenesis, the latter lead to a set of testable predictions on the same neutrino parameters currently under experimental investigations.

The emerging scenario selects the normal ordering of the neutrino mass pattern, a large value for the reactor mixing angle, $2\deg \lesssim \tre \lesssim 20\deg$, as well as a non maximal atmospheric mixing angle, $16\deg \lesssim \tatm \lesssim 41\deg$, and favours negative values for the Dirac phase \d. The signature of the proposed strong thermal \soten~solutions is in the relation obtained between the effective Majorana mass and the lightest neutrino mass: $\mee \approx 0.8 \, m_1 \approx 15 $ meV. 
\end{abstract}

\vspace{1 cm}

%===============================================================================
%-------------------------------------------------------------------------------
\section{Introduction} % (fold)
\label{sec:Introduction}
%-------------------------------------------------------------------------------
In the last decades, neutrino oscillation  experiments %\cite{Ahmad:2002jz,An:2012eh,Fukuda:1996sz}
 have underlined a fault in the current description of these particles. The detected squared mass splittings, not accounted for by the Standard Model (SM) of particle physics, pose the still open problem of neutrino mass origin. Beside that, during the same years, improved cosmological measurements
%\cite{Smoot:1992td,Komatsu:2011lr}
 revealed an asymmetry between the amount of matter and anti-matter in our Universe. A precise measurement of the former is given in the baryon asymmetry of the Universe (BAU)\cite{Ade:2013zuv}
\begin{equation}
	\label{eq:etab}
	\eta_B 
	\deq \frac{n_B - n_{\bar B}}{n_\gamma} 
	= (6.1 \pm 0.1) \times 10^{-10}
\end{equation}
which quantifies the overabundance of baryons over antibaryons with respect to the abundance of photons in our Universe. With no traces of primordial antimatter detected to date and with the reported value being too large to be explained by the SM physics, the BAU can thus be regarded as a further evidence in favour of physics beyond the current paradigms.

These problems, although apparently unrelated, find a common explanation in Leptogenesis\cite{Fukugita:1986hr}: a class of scenarios where the baryon asymmetry observed in our Universe is a natural consequence of the same mechanism behind the origin of neutrino masses. In its most minimal embodiment, Leptogenesis is based on the type-I Seesaw extension of the SM%\cite{GellMann:1980vs,Yanagida:1979as,1977PhLB...67..421M,1980PhRvL..44..912M}
, where three super-massive Majorana fermions are added as singlets to the particle content of the theory. These new particles also couple to the  left-handed lepton and Higgs doublets via a new set of Yukawa coupling $h$, effectively playing the role of right-handed neutrinos (RHN). Hence, after the Electro-Weak symmetry breaking, the Lagrangian for the considered model is
\begin{equation}
	\label{eq:lag}
	- {\LG} \supset  
	\overline{\alpha_L} \, D_{m_{\alpha}}\,\alpha_R 
	+ 
	\overline{\nu_{\alpha L}}\,m_{D\alpha i} \, N_{i R} 
	+
	\frac{1}{2} \, \overline{N^{c}_{i R}} \, {D_{M}}_i \, N_{i R}  
	+ \text{H.c.}; \qquad i\in\{1,2,3\}
\end{equation}
where $m_D = v h$, $v$ is the Higgs v.e.v., $\alpha$ are the charged leptons fields and $ D_{X} \deq \text{diag}(X)$, for instance $D_M \deq \text{diag}(M_1, M_2, M_3)$ with $M_1 \leq  M_2 \leq M_3$. 

By setting the RHN mass scale well above the Electro-Weak one, we can invoke the Seesaw mechanism as a solution to the neutrino mass puzzle. The neutrino spectrum then splits into two sets; the first one contains the particles $N_i \approx N_{iR} + N_{iR}^c$ whose masses almost coincide with the Majorana masses $M_i$. The remaining set contains three light Majorana neutrinos, which are responsible for the detected oscillation phenomena. The associated mass matrix
\begin{equation}
	\label{eq:seesaw}
	m_{\nu} = - m_D \, D_M^{-1} \, m_D^T; 
	\qquad 
	U^{\dagger} \,  m_{\nu} \, U^{\star}  =  - D_m. 
\end{equation}
is diagonalised by the PMNS mixing matrix $U$.

For the provided Yukawa couplings, the heavy RHN decay into lepton and Higgs doublets. In this regards, notice that the leptonic number is not conserved already at the Lagrangian level and that, in general, the interference between tree-level and 1-loop decay diagrams breaks the $CP$ symmetry. Hence, once the RHN decay out of equilibrium consequently to the Universe expansion, leptogenesis satisfies all the necessary conditions\cite{Sakharov:1967dj} to raise a lepton asymmetry in our Universe. The latter is then partially converted into the baryon asymmetry that we observe today by the sphaleron processes of the SM. By net leptogenesis therefore accounts for a BAU as large as $\eta_B^{lept} \approx 10^{-2} N_{B-L}^{lept}$, being $\lept$ the generated $B-L$ asymmetry abundance normalised to the abundance of  ultra-relativistic fermions in thermal equilibrium.       
% section Introduction (end)

%-------------------------------------------------------------------------------
\section{The problem of initial conditions} % (fold)
\label{sec:The problem of initial conditions}
%-------------------------------------------------------------------------------
Even though leptogenesis can generate the necessary BAU through the RHN decays, a priori we have no reasons to exclude the contributions of possible competing mechanisms into this quantity. In particular, for the high reheating temperature required by hierarchical thermal leptogenesis models, external mechanisms such as gravitational baryogenesis, 
%\cite{Davoudiasl:2004gf}
 Affleck-Dine baryogenesis 
 %\cite{Affleck:1984fy}  
 and even the traditional grand unified baryogenesis models, %\cite{doi:10.1146/annurev.ns.33.120183.003241} 
 might result in a sizeable contribution to the BAU before the onset of leptogenesis. This preexisting contribution, modelled in the preexisting $B-L$ asymmetry $\preex$, is in fact also recast as a baryon asymmetry by the SM sphalerons. For this reason it is not guaranteed that the observed BAU is a pure leptogenesis product and the value in Eq.~\eqref{eq:etab} cannot be employed to constrain the parameter space of a leptogenesis model. 

The \emph{problem of leptogenesis' initial conditions} posed by $\preex$ is elegantly solved by strong thermal leptogenesis (STL). STL is realised when the same processes that generate $\lept$ wash-out the preexisting component, in a way that at the end of the leptogenesis process $\lept \gg \preex$ and the leptogenesis contribution dominates the generated BAU. Within the simplified scenarios of $N_1$ leptogenesis STL can be achieved by simply requiring a strong washout regime%\cite{Buchmuller:2004nz}
. On the contrary, accounting for flavour effects\cite{JHEP092006010,Nardi:2006fx} complicates the problem to such an extent that successful STL is achieved only within the $\tau$\emph{on $N_2$-dominated scenario}\cite{Bertuzzo:2010fk}.  
More in detail, the latter is realised for an hierarchical RHN mass spectrum
\begin{equation}
	\label{eq:massRHN}
	M_1 \ll 10^9 \text{ GeV} < M_2 \leq 10^{12} \text{ GeV} \ll M_3
\end{equation}    
if particular conditions on the flavour decay-efficiency parameters $K_{i\alpha}$ (here and in the following $i~\in~\{1,2,3\}$, $\alpha\in\{e, \mu, \tau\}$) are satisfied:
\begin{equation}
	\label{eq:str_con}
	K_{1e}\gg 1, \quad K_{1\mu}\gg 1, \quad K_{1\tau}\lesssim 1,\quad K_{2\tau} > 1. 
\end{equation}
When the above requirements are satisfied, STL is guaranteed\cite{Bertuzzo:2010fk} as a consequence of the \emph{projection effect}\cite{JHEP092006010,Blanchet:2011xq}.   
% section The problem of initial conditions (end)

%-------------------------------------------------------------------------------
\section{The $SO(10)$-inspired model of leptogenesis} % (fold)
\label{sec:The $SO(10)$-inspired model of leptogenesis}
%-------------------------------------------------------------------------------
From the above discussion it should be clear that within the $\tau$on $N_2$-dominated scenario we can safely impose the bound in Eq.~\eqref{eq:etab} to effectively constrain the parameter space of the underlaying leptogenesis model.
However, with the Lagrangian in Eq.~\eqref{eq:lag} that introduces 18 new parameters in the model (15 entries in the neutrino Dirac mass matrix $m_D$ and the three RHN masses), the predictivity of the proposed framework might be doubted. The problem is solved by considering the following phenomenological parametrization. Introducing the bi-unitary decomposition of the Dirac mass matrix, $m_D = V_L^\dg D_{m_D} U_R$, it can be shown\cite{di-Bari:2009qy,DiBari:2010ux} that the unitary matrix $U_R$ is completely determined in the diagonalization of the matrix $M^{-1}(M^{-1})^\dagger$, where $M^{-1} = U_R D_M^{-1} U_R^T = D_{m_D}^{-1} V_L U D_m U^T V_L^T D_{m_D}^{-1}$ and the last step follows from Eq.~\eqref{eq:seesaw}. It is then possible to adopt the following quantities as input parameters of the model: three light neutrino masses $m_i$,  three mixing angles, two Majorana phases and one Dirac phase in the PMNS matrix $U$,  three Dirac neutrino masses $m_{D_i}$ and 3 mixing angles and 3 phases of the unitary matrix $V_L$. In this way, whereas part of the first nine parameters is effectively constrained by the neutrino oscillation experiments, it is possible to impose a theoretical scheme to bound the remaining quantities. We choose the \emph{$SO(10)$-inspired conditions}\cite{di-Bari:2009qy,DiBari:2010ux}, which drawing from $SO(10)$ grand-unified theories lead to the following requirements:
\begin{enumerate}[i)]
	\item Each neutrino Dirac mass $m_{D_i}$ is proportional to the one of the 
	up-type quark in the same SM generation: $m_{D_i} = \alpha_i 
	m_{u_i}$ and $\alpha_i = \mathcal{O}(1)$.
	\item As $V_L$ is the matrix that in the absence of the Seesaw mechanism would play the same role as the CKM matrix does within the quark sector, the mixing angles of the former are limited by the values of the counterparts in the latter.  
\end{enumerate}      
 
The $SO(10)$-inspired conditions define the theoretical framework adopted in our work: the \emph{$SO(10)$-inspired model of leptogenesis}. Through the Seesaw formula in Eq.~\eqref{eq:seesaw}, the requirement $i)$ leads in a natural way to the RHN mass pattern required by STL, Eq.~\eqref{eq:massRHN}. In this way the production of the $B-L$ asymmetry, $\lept$, proceeds through a sequence of distinguished stages that are described by simple sets of flavoured Boltzmann equations\cite{DiBari:2013qja,Blanchet:2011xq}. Furthermore, should the mentioned conditions on the decay efficiency parameters be satisfied, the $SO(10)$-inspired model would constitute a tangible framework for the realization of the $\tau$on $N_2$-dominated scenario.
% section The $SO(10)$-inspired model of leptogenesis (end)

%-------------------------------------------------------------------------------
\section{Strong thermal $SO(10)$-inspired leptogenesis} % (fold)
\label{sec:Results}
%-------------------------------------------------------------------------------
Investigating the compatibility between the $\tau$on $N_2$-dominated scenario and the $SO(10)$-inspired model is the subject of the highlighted study. We performed a thorough analysis of the parameter space of the model seeking for regions where on top of successful leptogenesis, \emph{i.e.} producing an amount of BAU compatible with the bounds of Eq.~\eqref{eq:etab}, also STL could be achieved. 
     
\begin{figure}[H]
	\centering
		\includegraphics[width=.47\textwidth]{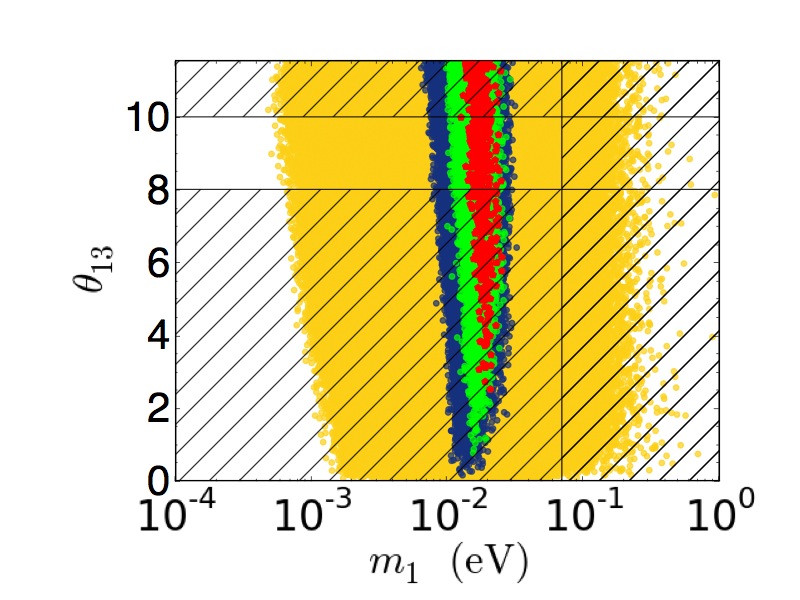}
		\includegraphics[width=.47\textwidth]{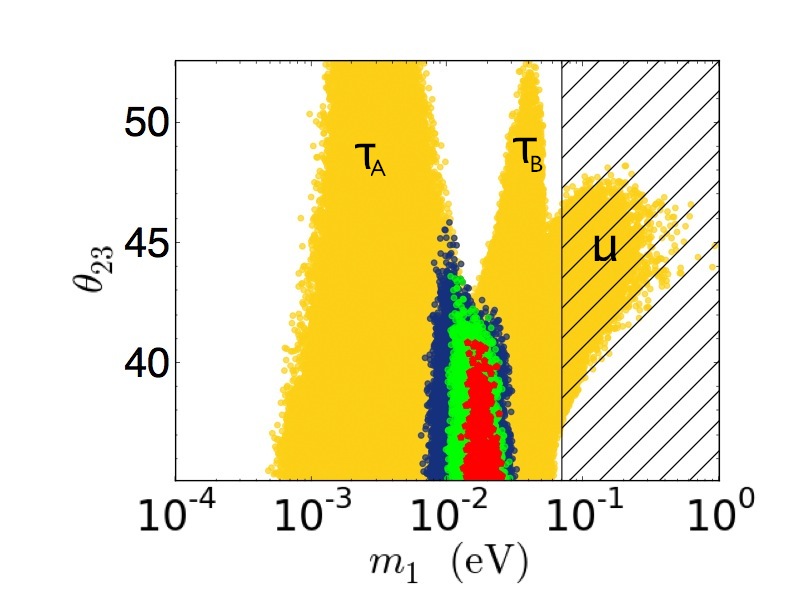}
		\\
		\includegraphics[width=.47\textwidth]{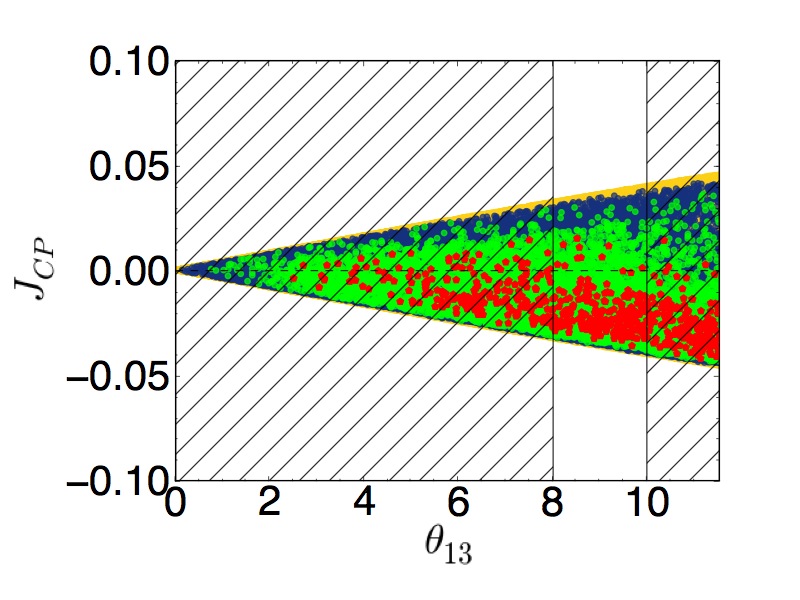}
		\includegraphics[width=.47\textwidth]{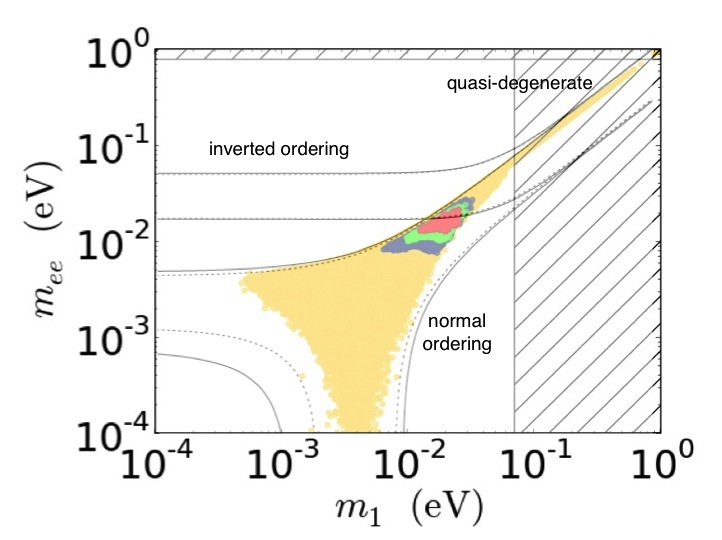}
	\caption{Successful strong thermal solution of the $SO(10)$-inspired model; normal ordering. In the yellow regions the model leads to an amount of baryon asymmetry of the Universe compatible with the bounds in Eq.~\eqref{eq:etab}. The remaining coloured areas mark the strong thermal leptogenesis solutions. Here, beside producing the required asymmetry, the leptogenesis process erases preexisting $B-L$ asymmetries as large as $\preex = 0.1$ (red), $ 0.01$ (green) and $ 0.001$ (blue). The labels $\tau_A$, $\tau_B$ and $\mu$ in the top right panel distinguish the three kind of solutions identified in literature\cite{DiBari:2013qja,di-Bari:2009qy,DiBari:2010ux}, while the hatched areas correspond to $2\sigma$ experimental exclusions\cite{Fogli:2012fk,GonzalezGarcia:2012sz,PhysRevD.86.073012}.}
	\label{fig:results}
\end{figure}
\FloatBarrier

A selection of the obtained results is proposed in Figure~\ref{fig:results}. Here the yellow regions denote the areas in the parameter space of the $SO(10)$-inspired model where successful leptogenesis is achieved: $\eta_B \approx 10^{-2} \lept$. Our analysis updates the one in\cite{di-Bari:2009qy,DiBari:2010ux}, confirming the lower bound on the lightest neutrino mass, $m_1$, as well as the existence of the three kinds of solutions, $\tau_A$, $\tau_B$ and $\mu$, therein proposed. The novelty of our work is in the remaining coloured regions, which denote the successful STL solutions for which $\lept \gg \preex$. For the combinations of parameters that fall in the red, green and blue areas, beside producing the desired amount of $B-L$ asymmetry, the model is able to wash-out preexisting asymmetries $\preex$ respectively as large as $10^{-1}$, $10^{-2}$ and $10^{-3}$. 
We remark that the presented results apply exclusively to normal ordering of the light neutrino spectrum: a first achievement of our investigation is \emph{excluding the inverted ordering} of the former on the basis of its incompatibility with the desired STL solutions.

As shown in Figure~\ref{fig:results}, the successful STL solution of the $SO(10)$-inspired model imply precise constraints on the neutrino oscillation parameter that can be regarded as predictions of our framework. From the top left panel it is clear that the proposed solutions support \emph{large values of the reactor mixing angle $\theta_{13}$}, plotted here against the lightest neutrino mass $m_1$. The bulk of solutions is found for $\theta_{13} \gtrsim 5 \degree$, well in agreement with current experimental results\cite{Fogli:2012fk,GonzalezGarcia:2012sz,PhysRevD.86.073012}. The top right panel shows instead that our scheme favourites \emph{a non maximal atmospheric mixing angle $\theta_{23}$}. Imposing the complete wash-out of $\preex = 10^{-1}$ $(10^{-3})$ yields a sharp upper bound $\theta_{23}\lesssim 41 \degree$ ($46 \degree$) which represents a distinguishing feature of our scenario. Interestingly, the quoted neutrino analyses favour a non-maximal atmospheric mixing angle in conjunction with a negative value of the Dirac phase $\delta$. As shown in the bottom left panel of Figure~\ref{fig:results}, this is also the case for our solutions. The Jarlskog invariant $J$, proportional to $\sin\delta$, is plotted here against the reactor mixing angle. Within the allowed $2\sigma$ range of the latter, the bulk of our solutions falls for $J<0$. Finally, the bottom right panel shows the neutrino effective Majorana mass $m_{ee}$, investigated by $0\nu\beta\beta$ decay experiments, as a function of the lightest neutrino mass. The successful STL solutions entangle the mentioned neutrino mass scales in the relation $\mee \approx 0.8 \, m_1 \approx 15 $ meV, which constitutes \emph{the signature} of the proposed scenario and can be tested by cosmological and next generation $0\nu\beta\beta$ experiments. 

We refer to\cite{DiBari:2013qja} for a complete overview of the results obtained with our method. A detailed study of the lower bound on $m_1$ in connection to the STL conditions has been proposed in\cite{DiBari:2014eqa}, while a dedicated investigation on the remaining distinguishing feature of the scenario is being finalised\cite{diBari20xx}.  

%-------------------------------------------------------------------------------
\section*{Acknowledgements} % (fold)
\label{sec:Acknowledgements}
%-------------------------------------------------------------------------------
I thank the organising committee for the kind invitation and the impeccable hospitality. I also thank Pasquale Di Bari and Michele Re Fiorentin for returning useful comments on the manuscript. My work is supported by the European Social Fund with the MOBILITAS grant MJD387.

% section Acknowledgements (end)

\bibliographystyle{hunsrt}
\bibliography{lm_proc}

\end{document}